\documentclass[10pt, final, conference, twocolumn]{IEEEtran/IEEEtran}
\usepackage{cite}
\usepackage{graphicx}
\usepackage{url}
\usepackage[centertags]{amsmath}
\usepackage{algorithm}
\usepackage{algorithmic}
\usepackage{amssymb}
\newcommand{\beq}{\begin{equation}}
\newcommand{\eeq}{\end{equation}}
\newcommand{\beqn}{\begin{align}}
\newcommand{\eeqn}{\end{align}}

\usepackage{mathtools}

\begin{document}
\IEEEspecialpapernotice{(Invited Paper)}
\title{Physical Layer Security for Massive MIMO Systems Impaired by Phase Noise}

\author{\IEEEauthorblockN{Jun~Zhu, Robert~Schober, and Vijay~K.~Bhargava}\\
\vspace{-3mm}
\IEEEauthorblockA{The University of British Columbia}
 \vspace*{-1cm}}
\IEEEoverridecommandlockouts

\setcounter{page}{1}

\maketitle

\begin{abstract}
In this paper, we investigate the impact of phase noise on the secrecy performance of downlink massive MIMO systems in the presence of a passive multiple-antenna eavesdropper. Thereby,
for the base station (BS) and the legitimate users, the effect of multiplicative phase noise is taken into account, whereas the eavesdropper is assumed to employ ideal hardware. We derive a lower bound for the ergodic secrecy rate of a given user when matched filter data precoding and artificial noise transmission are employed at the BS. Based on the derived
analytical expression, we investigate the impact of the various system parameters on the secrecy rate. Our analytical and simulation results reveal that distributively deployed local oscillators (LOs) can achieve a better performance than one common LO for all BS antennas as long as a sufficient amount of power is assigned for data transmission.
\end{abstract}
%%%%%%%%%%%%%%%%%%%%%%%%%%%%%%%%%%%%%%%%%%%%%%%%%%%%%%%%
%
%\begin{keywords}
%Physical layer security, artificial noise, massive MIMO, and phase noise.
%\end{keywords}
\section{Introduction}
%%%%%%%%%%%%%%%%%%%%%%%%%%%%%%%%%%%%%%%%%%%%%%%%%%%%%%%%
Massive multiple-input multiple-output (MIMO) systems promise tremendous performance gains in terms of network throughput and energy efficiency
by employing simple coherent processing across arrays of hundreds or even thousands of base station (BS) antennas, serving tens or hundreds of mobile terminals \cite{survey,noncooperative}. As an additional benefit, massive MIMO is inherently more
secure than conventional MIMO systems, as the large-scale antenna array equipped at the BS (Alice) can accurately focus a narrow and directional information beam on the intended terminals (Bob), such that the received signal power at Bob is several orders of magnitude higher than that at any incoherent passive eavesdropper (Eve) \cite{phymassive}. Unfortunately, this benefit may vanish if Eve also employs a massive antenna array for eavesdropping. In this case, unless additional measures to secure the communication are taken by Alice, even a single passive Eve is able to intercept the signal intended for Bob \cite{zhu}.

Since security is a critical concern for future communication systems, facilitating secrecy at the physical layer of massive MIMO systems has received significant
attention recently. Artificial noise (AN) generation \cite{negi} was employed to provide physical
layer security in a multi-cell massive MIMO system with pilot contamination in \cite{zhu}. Thereby, it was shown that secure communication can be achieved even with
simple matched filter (MF) precoding of the data and null-space (NS) precoding of the AN.  Nevertheless, it was revealed in \cite{zhu2} that significant additional
performance gains are possible with more sophisticated data and AN precoders, including polynomial precoders. Furthermore, AN-aided jamming of Rician fading massive MIMO
channels was investigated in \cite{jwang}.

All aforementioned works on secure massive MIMO are based on the assumption that the transceivers of the legitimate users are equipped with perfect hardware components, i.e.,
the effects of hardware impairments were not taken into account. Nevertheless, all practical implementations do suffer from impairments arising from non-ideal hardware components \cite{nonideal}. These impairments are expected to be particularly pronounced in massive MIMO systems as the excessive number of
BS antennas makes the use of low-cost components desirable to keep the overall capital expenditures for operators manageable. Several works
have investigated the impact of hardware impairments on massive MIMO systems \cite{nonideal}-\cite{TRMRC}. These works demonstrated that hardware impairments can severely limit the performance of massive MIMO
systems, and that phase noise originating from free-running oscillators is the main contributor in degrading the quality of the channel state information (CSI) estimates needed for precoder design. On the one hand, phase noise
causes the CSI estimates to become outdated more quickly, and on the other hand, it may cause a loss of orthogonality of the pilot sequences employed by the different users in a cell for uplink training. If communication secrecy is considered, an additional challenge arises: Whereas the legitimate users of the system will likely employ
low-cost equipment giving rise to hardware impairments, the eavesdropper is expected to employ high-quality equipment and avoid hardware impairments. This disparity in equipment quality was not considered in the related work on physical layer security \cite{zhu}-\cite{jwang}
nor in the related work on hardware impairments \cite{nonideal}-\cite{TRMRC} and necessitates the development of a new analysis and design framework.

Motivated by the above considerations, in this paper, we present the first study of physical layer security for massive MIMO systems in the presence of phase noise. Thereby, we focus on the
downlink and investigate the effects of multiplicative phase noise at the BS and the users. As a worst-case scenario, the eavesdropper is assumed to employ ideal (phase noise-free) hardware. We derive a tight lower bound for the ergodic secrecy rate achieved by a downlink user in the presence of phase noise when MF data precoding and NS AN precoding
are employed at the massive MIMO BS. The derived bound provides insight into the impact of various system and channel parameters, such as the phase noise variance, the amount of power allocated to the AN, the number of users, and the number of deployed local oscillators (LOs) on the ergodic secrecy rate.

\textit{Notation:} Superscripts $T$ and $H$ stand for the transpose and conjugate transpose, respectively. ${\bf I}_N$ is the $N$-dimensional identity matrix. The expectation operation of a random variable is denoted by $\mathbb{E}[\cdot]$. ${\rm diag}\{{\bf x}\}$ denotes a diagonal matrix with the elements of vector
${\bf x}$ on the main diagonal. $\mathbb{C}^{m\times n}$ represents the space of all $m \times n$ matrices with complex-valued elements.
${\bf x} \sim \mathbb{CN}({\bf 0}_N, \boldsymbol{\Sigma})$ denotes a circularly symmetric complex Gaussian vector ${\bf x}\in \mathbb{C}^{N \times 1}$ with zero mean and covariance
matrix $\boldsymbol{\Sigma}$. $[{\bf A}]_{kl}$ denotes the element in the $k^{\rm th}$ row and $l^{\rm th}$ column of matrix ${\bf A}$. Finally, $[x]^+=\max\{x,0\}$.
%%%%%%%%%%%%%%%%%%%%%%%%%%%%%%%%%%%%%%%%%%%%%%%%%%%%%%%%%%%%%%%%%%%%%%%%%%%%%%%%%%%
%%%%%%%%%%%%%%%%%%%%%%%%%%%%%%%%%%%%%%%%%%%%%%%%%%%%%%%%%%%%%%%%%%%%%%%%%%%%%%%%%%%
\section{System and Channel Models}\label{s2}
%%%%%%%%%%%%%%%%%%%%%%%%%%%%%%%%%%%%%%%%%%%%%%%%%%%%%%%%%%%%%%%%%%%%%%%%%%%%%%%%%%%
The considered massive MIMO system model comprises an $N$-antenna BS, $K$ single-antenna mobile terminals (MTs), and an $N_E$-antenna eavesdropper. The eavesdropper is passive
in order to hide its existence from the BS and the MTs. Similar to \cite{pn,TRMRC}, we assume that after proper compensation the residual hardware impairments manifest themselves at the
BS and the MTs in the form of multiplicative phase noise. The impact of the phase noise on uplink training and downlink data transmission is investigated in Sections \ref{s2a} and
\ref{s2b}, respectively, and the signal model for the eavesdropper is presented in Section \ref{s2c}.
%%%%%%%%%%%%%%%%%%%%%%%%%%%%%%%%%%%%%%%%%%%%%%%%%%%%%%%%%%%%%%%%%%%%%%%%%%%%%
\subsection{Uplink Pilot Training under Phase Noise}\label{s2a}
%%%%%%%%%%%%%%%%%%%%%%%%%%%%%%%%%%%%%%%%%%%%%%%%%%%%%%%%%%%%%%%%%%%%%%%%%%%%%%%
%%%%%%%%%%%%%%%%%%%%%%%%%%%%%%%%%%%%%%%%%%%%%%%%%%%
In massive MIMO systems, the CSI is usually acquired via uplink training by exploiting the channel reciprocity between uplink and downlink \cite{noncooperative}. As such, in the training phase (the beginning of each coherence block of length $T$), all MTs emit mutually orthogonal pilot sequences $\boldsymbol{\omega}_k=[\omega_k(1),\omega_k(2),\ldots,\omega_k(B)]^T \in \mathbb{C}^{B \times 1},1 \leq k\leq K$ of length $B$ ($B \geq K$). We have $\boldsymbol{\omega}^H_k \boldsymbol{\omega}_{k'}=B p_\tau$ for $k'=k$, and $\boldsymbol{\omega}^H_k \boldsymbol{\omega}_{k'}=0$ otherwise. The received vector ${\bf y}_{\rm tr}(t) \in \mathbb{C}^{N \times 1},1 \leq t \leq B$, at the BS is given by
\begin{equation}\label{yULt}
{\bf y}_{\rm tr}(t)=\sum_{k=1}^K \boldsymbol{\Theta}_k(t){\bf g}_k \omega_k(t)+\boldsymbol{\xi}^{\rm UL}(t).
\end{equation}
Here, the channel vector of the $k^{\rm th}$ MT, ${\bf g}_k \sim \mathbb{CN}({\bf 0}_N, \beta_k{\bf I}_N)$, is modelled as block Rayleigh fading, where $\beta_k$ denotes the
path-loss. Thereby, ${\bf g}_k$ is assumed to be constant during coherence time $T$ and change independently afterwards. $\boldsymbol{\xi}^{\rm UL}(t) \sim \mathbb{CN}(0,\xi^{\rm UL})$ denotes the thermal noise at the BS. In (\ref{yULt}), the term
$\boldsymbol{\Theta}_k(t)$ characterizes the phase noise affecting the uplink
training, and is given by $\boldsymbol{\Theta}_k(t)={\rm diag}\left(e^{j \theta^1_k(t)} {\bf 1}_{1 \times N/N_o},\ldots,e^{j \theta^{N_o}_k(t)}{\bf 1}_{1 \times N/N_o}\right) \in \mathbb{C}^{N \times N}$, where $N_o$ denotes the number of free-running LOs equipped at the BS \cite{pn}. Thereby, we assume that at the BS each group of $N/N_o \in \mathbb{Z}$
antennas is connected to one free-running LO. $\theta^l_k(t)=\psi_l(t)+\phi_k(t)$ is the phase noise that affects the link between the $l^{\rm th}$ LO at the BS and the
$k^{\rm th}$ MT. Adopting the discrete-time Wiener phase noise model \cite{pn}, in time interval $t$, the phase noises at the $l^{\rm th}$ LO of the BS and the $k^{\rm th}$ MT
are modelled as $\psi_l(t) \sim \mathbb{CN}(\psi_l(t-1), \sigma^2_{\psi})$, $1 \leq l \leq N_o$, and $\phi_k(t) \sim \mathbb{CN}(\phi_k(t-1), \sigma^2_{\phi})$, $1 \leq k \leq K$, where
$\sigma^2_{\psi}$ and $\sigma^2_{\phi}$ are the phase noise (increment) variances at the BS and the MTs, respectively.\\

For channel estimation, we collect the signal vectors received during the training phase in vector $\boldsymbol{\psi}=[{\bf y}^T_{\rm tr}(1),\ldots,{\bf y}^{T}_{\rm tr}(B)]^T \in \mathbb{C}^{BN \times 1}$, and define the effective channel vector at time $t$ as ${\bf g}_k(t)=\boldsymbol{\Theta}_k(t){\bf g}_k$. With these definitions,
the linear minimum mean-square error (LMMSE) estimate of the channel of the $k^{\rm th}$ MT at time $t$ (i.e., during the data transmission phase) can be written as \cite{MRC}
\begin{equation}\label{hatgkt1}
\hat{\bf g}_k(t)=\left(\beta_k \boldsymbol{\omega}^H_k \boldsymbol{\Theta}_{\sigma(t)} \boldsymbol{\Sigma}^{-1} \otimes {\bf I}_N\right) \boldsymbol{\psi},
\end{equation}
where
\begin{equation}\label{thetasigma}
\boldsymbol{\Theta}_{\sigma(t)}={\rm diag}\left(e^{-\frac{\sigma^2_\psi+\sigma^2_\phi}{2}|t-1|},\ldots,e^{-\frac{\sigma^2_\psi+\sigma^2_\phi}{2}|t-B|}\right) \end{equation}
and $\boldsymbol{\Sigma}=\sum_{k=1}^K \beta_k {\bf W}_k+\xi^{\rm UL} {\bf I}_{B}$ with $[{\bf W}_k^b]_{i,j}=\omega_k(i)\omega^*_k(j) e^{-\frac{\sigma^2_\psi+\sigma^2_\phi}{2}|i-j|}$, $i,j \in \{1,\ldots B\}$.

Considering the properties of LMMSE estimation, the channel can be decomposed as ${\bf g}_k(t)=\hat{\bf g}_k(t)+{\bf e}(t)$, $t=1,\ldots,B$, where $\hat{\bf g}_k(t)$ denotes the
LMMSE channel estimate given in (\ref{hatgkt1}) and ${\bf e}_k(t)$ represents the estimation error. $\hat{\bf g}_k(t)$ and ${\bf e}(t)$ are mutually uncorrelated and have zero mean
\cite[{Theorem 1}]{nonideal}. The error covariance matrix is given by
\begin{equation}\label{error}
\mathbb{E}[{\bf e}_k(t){\bf e}^H_k(t)]=\beta_k \left(1-\beta_k \boldsymbol{\omega}_k^H \boldsymbol{\Theta}_{\sigma(t)} \boldsymbol{\Sigma}^{-1} \boldsymbol{\Theta}_{\sigma(t)} \boldsymbol{\omega}_k\right) {\bf I}_N.
\end{equation}
Eqs.~(\ref{yULt})-(\ref{error}) reveal that, in general, for $\sigma^2_\psi, \sigma^2_\phi >0$, the channel estimate of the $k^{\rm th}$ MT contains contributions from channels of other MTs, i.e., pilot contamination occurs although the emitted pilots are orthogonal. Furthermore, (\ref{hatgkt1}) reveals that the channel estimate depends on time $t$. As a consequence, ideally, the channel-dependent data and AN precoders employed for downlink transmission
should be recomputed for every symbol interval of the data transmission phase, which entails a high computational complexity. Therefore, in the following, we assume that the data and AN precoders are computed
based on the channel estimate for one symbol interval $t_0$ (e.g.~$t_0=B+1$) and are then employed for precoding during the entire data transmission phase, i.e., for $t \in \{B+1,\ldots,T\}$. For notational conciseness,
we denote the corresponding channel estimate by $\hat{\bf g}_{k}=\hat{\bf g}_{k}(t_0)$, $k= \{1,\ldots K\}$.
%%%%%%%%%%%%%%%%%%%%%%%%%%%%%%%%%%%%%%%%%%%%%%%%%%%%%%%%%%%%%%%%%%%%%%%%%
\subsection{Downlink Data Transmission and Linear Precoding}\label{s2b}
%%%%%%%%%%%%%%%%%%%%%%%%%%%%%%%%%%%%%%%%%%%%%%%%%%%%%%%%%%%%%%%%%%%%%%%%
Assuming channel reciprocity, during the downlink data transmission phase, the received signal at the $k^{\rm th}$ MT in time interval $t \in \{B+1,\ldots,T\}$ is given by
\begin{equation}\label{ykt}
y^{\rm DL}_k(t)={\bf g}^H_k \boldsymbol{\Theta}^H_k(t) {\bf x}+\xi^{\rm DL}_k(t),
\end{equation}
where $\xi^{\rm DL}_k(t) \sim \mathbb{CN}(0,\xi^{\rm DL})$ represents the thermal noise at the $k^{\rm th}$ MT. The downlink transmit signal ${\bf x} \in \mathbb{C}^{N \times 1}$ in (\ref{ykt}) is modeled as
\begin{equation}\label{x}
{\bf x}=\sqrt{p}{\bf F}{\bf s}+\sqrt{q}{\bf A}{\bf z} \in \mathbb{C}^{N \times 1},
\end{equation}
where the data symbol vector ${\bf s} \in \mathbb{C}^{K \times 1}$ and the AN vector ${\bf z} \in \mathbb{C}^{L \times 1}$, $L \leq N$, are multiplied by data precoder ${\bf F} \in \mathbb{C}^{N \times K}$  and AN
precoder ${\bf A} \in \mathbb{C}^{N \times L}$, respectively. As we assume that the eavesdropper's CSI is not available at the BS, AN is injected to degrade the eavesdropper's ability to decode the data
intended for the MTs \cite{zhu,zhu2,zhou}. Thereby, it is assumed that the components of ${\bf s}$ and ${\bf z}$ are independent and identically distributed (i.i.d.) circularly symmetric complex Gaussian (CSCG)
random variables, i.e., ${\bf s} \sim \mathbb{CN}({\bf 0}_K,{\bf I}_K)$ and ${\bf z} \sim \mathbb{CN}({\bf 0}_L,{\bf I}_L)$. In (\ref{x}), $p=\phi P_T/K$ and $q=(1-\phi) P_T/L$ denote the power assigned to each
MT and each column of the AN, where $P_T$ is the total power budget and $\phi \in (0,1]$ can be used to strike a balance between data transmission and AN emission. Combining (\ref{x}) and (\ref{ykt}) we obtain $y^{\rm DL}_k(t)=$
\begin{equation}\label{yDLkt}
\sqrt{p}{\bf g}^H_k(t) {\bf f}_k s_k+\sum_{l \neq k}^K \sqrt{p}{\bf g}^H_k(t) {\bf f}_l s_l +\sqrt{q} {\bf g}^H_k(t) {\bf A}{\bf z}+\xi^{\rm DL}_k(t),
\end{equation}
where $s_k$ and ${\bf f}_k$ denote the $k^{\rm th}$ element of ${\bf s}$ and the $k^{\rm th}$ column of matrix ${\bf F}$, respectively.
%%%%%%%%%%%%%%%%%%%%%%%%%%%%%%%%%%%%%%%%%%%%%%%%%%%%%%%%%
\subsection{Signal Model of the Eavesdropper}\label{s2c}
%%%%%%%%%%%%%%%%%%%%%%%%%%%%%%%%%%%%%%%%%%%%%%%%%%%%%%%%%%%%
We assume that the eavesdropper is silent during the training phase, i.e., for $t \in \{1,\ldots,B\}$, and eavesdrops the signal intended for the $k^{\rm th}$ MT during the
data transmission phase, i.e., for $t \in \{B+1,\ldots,T\}$. Let ${\bf G}_E$ denote the channel matrix between the BS and the eavesdropper with i.i.d.~zero-mean complex Gaussian elements having variance $\beta_E$,
where $\beta_E$ is the path-loss between the BS and the eavesdropper. Since the capabilities of the eavesdropper are not known at the BS, we make worst-case assumptions regarding the hardware and signal
processing capabilities of the eavesdropper with respect to communication secrecy. In particular, we assume the received signal at the eavesdropper at time $t \in \{B+1,\ldots,T\}$ can be modelled as
\begin{equation}\label{yEt}
{\bf y}_E(t)={\bf G}_E^H \boldsymbol{\Psi}^H(t) {\bf x} \in \mathbb{C}^{N_E \times 1},
\end{equation}
where $\boldsymbol{\Psi}(t)={\rm diag}\left(e^{j \psi_1(t)} {\bf 1}^T_{1 \times N/N_o},\ldots,e^{j \psi_{N_o}(t)}{\bf 1}^T_{1 \times N/N_o}\right)$. Thereby, we assumed that the eavesdropper employs high-quality
hardware such that the only hardware impairment is the phase noise at the BS. Eq.~(\ref{yEt}) also implies that the thermal noise at the eavesdropper is
negligibly small \cite{zhu,zhu2,zhou}. Furthermore, we assume that the eavesdropper has perfect CSI, i.e., it perfectly knows the effective eavesdropper channel matrix ${\bf G}_E^H\boldsymbol{\Psi}^H(t) $. We also assume that it can
can perfectly decode and cancel the interference caused by all MTs except for the MT of interest \cite{zhu,zhu2,zhou}. These worst-case assumptions lead to an upper bound on the ergodic capacity of the
eavesdropper given by
\begin{equation}\label{CE}
C_E=\mathbb{E}[\log_2(1+\gamma_E)],~\gamma_E=p {\bf g}^k_E \left(q{\bf G}_E^H {\bf A}{\bf A}^H {\bf G}_E\right)^{-1}({\bf g}^k_E)^H,
\end{equation}
where ${\bf g}^k_E={\bf f}_k^H {\bf G}_E$. We note that since we assumed that the thermal noise at the receiver of the eavesdropper is negligible, $\gamma_E$, and consequently $C_E$, are independent of the path-loss
of the eavesdropper, $\beta_E$. Furthermore, since perfect channel estimation at the eavesdropper was assumed, the impact of the phase noise is completely removed.
%%%%%%%%%%%%%%%%%%%%%%%%%%%%%%%%%%%%%%%%%%%%%%%%%%%%%%%%%%%%%%%%%%%
%%%%%%%%%%%%%%%%%%%%%%%%%%%%%%%%%%%%%%%%%%%%%%%%%%%%%%%%%%%%%%%%%%%%
\section{Achievable Ergodic Secrecy Rate in the Presence of Phase Noise}\label{s3}
%%%%%%%%%%%%%%%%%%%%%%%%%%%%%%%%%%%%%%%%%%%%%%%%%%%%%%%%%%%%%%%%%%%%%%
In this section, we analyze the achievable ergodic secrecy rate of  a massive MIMO system in the presence of phase noise. To this end, we derive a lower bound on the achievable ergodic
secrecy rate in Section \ref{s3a}, and present an asymptotic analysis for the downlink data rate of the legitimate MTs when MF data precoding and NS AN precoding are adopted by the BS in Section \ref{s3b}. In Section \ref{s3c}, a simple closed-form upper bound for the eavesdropper's capacity is presented.

%%%%%%%%%%%%%%%%%%%%%%%%%%%%%%%%%%%%%%%%%%%%%%%%%%%%%%%%%%%%%%%%%%%%%%
\subsection{Lower Bound on Achievable Ergodic Secrecy Rate}\label{s3a}
%%%%%%%%%%%%%%%%%%%%%%%%%%%%%%%%%%%%%%%%%%%%%%%%%%%%%%%%%%%%%%%%%%%%%%
Before analyzing the secrecy rate, we first employ \cite[{Lemma 1}]{MRC} to obtain a lower bound on the achievable rate of the multiple-input single-output (MISO) phase noise channel given by (\ref{ykt}).
In particular, the achievable rate of the $k^{\rm th}$ MT, $1 \leq k \leq K$, in symbol interval $t \in \{B+1,\ldots,T\}$ is lower bounded by
\begin{equation}\label{Rkt}
R_k(t) \geq \underline{R}_k(t)=\log_2 (1+\gamma_k(t)),
\end{equation}
with signal-to-interference-plus-noise ratio (SINR) $\gamma_k(t)$ given in (\ref{gammak}) at the top of the next page.
\begin{figure*}[!t]
\begin{equation}\label{gammak}
\gamma_k(t)=\frac{p\left|\mathbb{E}\left[{\bf g}^H_k(t) {\bf f}_k\right]\right|^2}{\sum\limits_{l=1}^K p\mathbb{E}\left[\left|{\bf g}^H_k(t) {\bf f}_l\right|^2\right]-p\left|\mathbb{E}\left[{\bf g}^H_k(t) {\bf f}_k\right]\right|^2
+q\mathbb{E}\left[{\bf g}^H_k(t) {\bf A}{\bf A}^H {\bf g}_k(t)\right]+\xi^{\rm DL}}
\end{equation}
\hrule
\end{figure*}
The expectation operator in (\ref{gammak}) is taken with respect to channel vectors, ${\bf g}_k$, as well as the phase noise processes, $\psi_l(t)$ and $\phi_k(t)$. The SINR in (\ref{gammak}) is
obtained by employing the average effective channel gain $\left|\mathbb{E}\left[{\bf g}^H_k(t) {\bf f}_k\right]\right|$ for signal detection, while treating the deviation from the average effective
channel gain as Gaussian noise having variance $\mathbb{E}\left[\left|{\bf g}^H_k(t) {\bf f}_k\right|^2\right]-|\mathbb{E}\left[{\bf g}^H_k(t) {\bf f}_k\right]|^2$. Moreover, following
\cite[Lemma 1]{MRC}, we treated the multiuser interference as independent Gaussian noise, which is a worst-case assumption for the calculation of the mutual information.
Based on (\ref{Rkt}), we provide a lower bound on the achievable ergodic secrecy rate of the $k^{\rm th}$ MT, $1 \leq k \leq K$, in the following Lemma.

\textit{Lemma 1}: The achievable ergodic secrecy rate of the $k^{\rm th}$ MT, $1\leq k \leq K$, is bounded below by
\begin{equation}\label{rseck}
{R}^{\rm sec}_k \geq \underline{R}^{\rm sec}_k =\frac{1}{T} \sum_{t \in \{B+1,\ldots,T\}} \left[\underline{R}_k(t)-C_E\right]^+,
\end{equation}
where $\underline{R}_k(t)$, $1 \leq k \leq K$, is the lower bound of the achievable ergodic rate of the $k^{\rm th}$ MT given in (\ref{Rkt}) and $C_E$ is the ergodic capacity between
the BS and the eavesdropper given in (\ref{CE}).
\begin{IEEEproof}
The ergodic secrecy rate achieved by the $k^{\rm th}$ MT in symbol interval $t \in \{B+1,\ldots,T\}$ is given by \cite[{Lemma 1}]{zhu}
\begin{eqnarray}
\nonumber R^{\rm sec}_k(t)&=&\mathbb{E}\left[[R_k(t)-\log_2(1+\gamma_E)]^+\right]{\geq} \left[\mathbb{E}[R_k(t)]-C_E\right]^+ \\
&\overset{(a)}{\geq}& \left[\underline{R}_k(t)-C_E\right]^+
=\underline{R}^{\rm sec}_k(t),
\end{eqnarray}
where $\underline{R}^{\rm sec}_k(t)$ is an achievable lower bound for $R^{\rm sec}_k(t)$, and $(a)$ uses (\ref{Rkt}). By averaging $R^{\rm sec}_k(t)$ over
all symbol intervals $t \in \{B+1,\ldots,T\}$ we obtain Lemma 1. This completes the proof.
\end{IEEEproof}
$C_E$ in (\ref{rseck}) is constant for all $t \in \{B+1,\ldots,T\}$ as we made the worst-case assumptions that the eavesdropper employs ideal hardware and has perfect CSI. The sum in (\ref{rseck})
is over the $T-B$ time slots used for data transmission. Motivated by the coding scheme for the non-secrecy case in \cite{TRMRC}, a similar coding scheme that supports the secrecy rate given in (\ref{rseck}) can be described as follows. For a given $t \in \{B+1,\ldots,T\}$, the statistics of ${\bf g}_k(t)$ in (\ref{gammak}) given the estimate $\hat{\bf g}_k$ are identical across all coherence intervals and the corresponding
channel realizations are i.i.d. Hence, we employ $T-B$ parallel channel codes for each MT; one code for each time $t \in \{B+1,\ldots,T\}$, i.e., the $t^{\rm th}$ channel code is employed across the $t^{\rm th}$ time slots  of multiple
coherence intervals. Then, at each MT, the $t^{\rm th}$ received symbols across the multiple coherence intervals are jointly decoded \cite{TRMRC}. With this coding strategy the ergodic secrecy rate given
in (\ref{rseck}) is achieved provided the parallel codes span sufficiently many (ideally an infinite number) of independent channel realizations ${\bf g}_k$  and phase noise samples $\psi_l(t)$ and $\phi_k(t)$.
%%%%%%%%%%%%%%%%%%%%%%%%%%%%%%%%%%%%%%%%%%%%%%%%%%%%%%%%%%%%%%%%%%%%%%
\subsection{Asymptotic Analysis of Achievable Rate for MF Precoding}\label{s3b}
%%%%%%%%%%%%%%%%%%%%%%%%%%%%%%%%%%%%%%%%%%%%%%%%%%%%%%%%%%%%%%%%%%%%%%
In this subsection, we analyze the lower bound on the achievable ergodic rate of the $k^{\rm th}$ MT, $1 \leq k \leq K$, in (\ref{Rkt}) in the asymptotic limit $N,K\to\infty$ for fixed ratio $\beta=K/N$. Thereby, we adopt
MF precoding at the BS, i.e., ${\bf f}_k=\hat{\bf g}_{k}/\|\hat{\bf g}_{k}\|$, as is commonly done for massive MIMO systems because of the high complexity of more sophisticated precoder designs. In the following Lemma,
we provide a closed-form expression for the gain of the desired signal.

\textit{Lemma 2}: For MF precoding at the BS, the numerator of (\ref{gammak}) reflecting the gain of the desired signal at the $k^{\rm th}$ MT, can be expressed as
\begin{equation}\label{lambdak}
 \mathbb{E}\left[{\bf g}^H_k \boldsymbol{\Theta}^H_k(t) {\bf f}_k\right]=\sqrt{\beta_k N \lambda_k} \cdot e^{-\frac{\sigma^2_\psi+\sigma^2_\phi}{2}|t-t_0|},
\end{equation}
where $\lambda_k=\beta_k \boldsymbol{\omega}_k^H \boldsymbol{\Theta}_{\sigma(t_0)} \boldsymbol{\Sigma}^{-1} \boldsymbol{\Theta}_{\sigma(t_0)} \boldsymbol{\omega}_k$.
\begin{IEEEproof}
Please refer to journal version \cite[Appendix B]{zhu3}.
\end{IEEEproof}
The term $e^{-\frac{\sigma^2_\psi+\sigma^2_\phi}{2}|t-t_0|}$ in (\ref{lambdak}) reveals the impact of the accumulated phase noise from the time of channel estimation, $t_0$, to the time of
data transmission, $t$, on the received signal strength at MT $k$. On the other hand, the phase noise within the training phase affects $\lambda_k$, and consequently the received signal strength, via
$\boldsymbol{\Theta}_{\sigma(t_0)}$ and $\boldsymbol{\Sigma}$, cf.~(\ref{thetasigma}), when multiple pilot sequences are simultaneously emitted.
Next, an expression for the multiuser interference power in the first term of the denominator of (\ref{gammak}) is derived.

\textit{Lemma 3}: When MF precoding is adopted at the BS, the power of the multiuser interference caused by the signal intended for the $l^{\rm th}$ MT, $l \neq k$,
is given by
\begin{equation}\label{ak1}
\mathbb{E}\left[\left|{\bf g}^H_k \boldsymbol{\Theta}^H_k(t) {\bf f}_l\right|^2\right]=\left(\beta_k+ \left(X^{(1)}_{k,l}+X^{(2)}_{k,l}\right)\left(\frac{1-\epsilon}{N_o}+\epsilon\right)\right), \end{equation}
where $\epsilon=e^{-\sigma^2_\psi | t-t_0|}$,  $X^{(1)}_{k,l}=\frac{N}{N_o} \times \frac{\beta^2_k \boldsymbol{\omega}^H_l \boldsymbol{\Theta}_{\sigma(t_0)}\boldsymbol{\Sigma}^{-1}{\bf W}_k\boldsymbol{\Sigma}^{-1}
\boldsymbol{\Theta}_{\sigma(t_0)}\boldsymbol{\omega}_l}{\boldsymbol{\omega}_l^H \boldsymbol{\Theta}_{\sigma(t_0)} \boldsymbol{\Sigma}^{-1}
\boldsymbol{\Theta}_{\sigma(t_0)} \boldsymbol{\omega}_l}$ and $X^{(2)}_{k,l}=N \left(1-\frac{1}{N_o}\right) \times \frac{\bigg|\beta_k \boldsymbol{\omega}_k^H
\boldsymbol{\Theta}_{\sigma(t_0)} \boldsymbol{\Sigma}^{-1} \boldsymbol{\Theta}_{\sigma(t_0)} \boldsymbol{\omega}_l\bigg|^2}{\boldsymbol{\omega}_l^H
\boldsymbol{\Theta}_{\sigma(t_0)} \boldsymbol{\Sigma}^{-1} \boldsymbol{\Theta}_{\sigma(t_0)} \boldsymbol{\omega}_l}$.
\begin{IEEEproof}
Please refer to journal version \cite[Appendix C]{zhu3}.
\end{IEEEproof}
\textit{Remark 1}: Lemma 3 reveals that when the number of BS antennas is sufficiently large, i.e., $N \to \infty$, the impact of the multiuser interference from the $l^{\rm th}$ MT grows linearly with $N$ and does
not vanish compared to the strength of the desired signal in (\ref{lambdak}) in the limit of $N\to\infty$ due to the impairment incurred by the phase noise during the training phase. Furthermore, although not obvious, it can be verified that the multiuser interference decreases with increasing number of LOs, $N_o$.

Furthermore, for the summand with $l=k$ in the sum in the first term of the denominator of (\ref{gammak}), we obtain
\begin{equation}\label{exs}
\mathbb{E}\left[\left|{\bf g}^H_k \boldsymbol{\Theta}^H_k(t) {\bf f}_k\right|^2\right]=\beta_k +\beta_k (N-1) \lambda_k \left(\frac{1-\epsilon}{N_o}+\epsilon\right),
\end{equation}
by applying \cite[Theorem 1]{zhu3} from free probability theory \cite{free}. The variance of the gain of the desired signal, ${\bf g}^H_k \boldsymbol{\Theta}^H_k(t) {\bf f}_k$, is obtained by subtracting the right hand side
of (\ref{exs}) from the square of the right hand side of (\ref{lambdak}).

Finally, the AN leakage incurred by the NS AN precoder is given in the following Lemma.

\textit{Lemma 4}: For the NS AN precoder, where $L=N-K$ \cite{zhu,zhu2,zhou}, the AN leakage power received at the $k^{\rm th}$ MT in time interval $t$ is given by
\begin{equation}\label{AN1}
L^k_{\rm AN}=\beta_k (N-K)\left(\left(1-\frac{1}{N_o}\right)\left(1-\epsilon\right)+1-\lambda_k\right).
\end{equation}
\begin{IEEEproof}
Please refer to journal version \cite[Appendix D]{zhu3}.
\end{IEEEproof}
\textit{Remark 2}: In Lemma 4, the terms $\epsilon$ and $\lambda_k$ reflect the negative impact of the hardware impairments on the AN power leakage. If a common LO is employed, i.e., $N_o=1$, the impact of $\epsilon$ is eliminated.
However, the negative effect of $\epsilon$ increases as the number of LOs, $N_o$, increases since the phase noise processes of different LOs are independent destroying the orthogonality of the columns of
${\bf A}$ and ${\bf g}_{k}(t)$, $1 \leq k \leq K$. Nevertheless, a common LO does not necessarily result in the maximum secrecy rate, as the multiuser interference decreases with increasing $N_o$, cf. Remark 1.

Substituting the results in (\ref{lambdak})-(\ref{AN1}) into (\ref{gammak}), the achievable rate of MT $k$ in time slot $t$ with MF data and NS AN precoding is obtained as
\begin{equation}\label{Rkunder}
\underline{R}_k(t) =\log_2 \left(1+\frac{\overline{\lambda}_k \phi N}{ (a_k+c_k-\beta \mu_k) \phi + \beta \mu_k +\xi_k}\right),
\end{equation}
where
\begin{equation}\label{ak}
a_k=\sum_{l \neq k}^K\left(1+ \left(X^{(1)}_{k,l}+X^{(2)}_{k,l}\right)\left(\frac{1-\epsilon}{N_o}+\epsilon\right)/\beta_k\right),
\end{equation}
\begin{equation}\label{ck}
c_k=\left(1-\frac{1}{N_o}\right)(1-\epsilon)+[(N-1) \lambda_k+1] \left(\frac{1-\epsilon}{N_o}
+ \epsilon\right)-N \overline{\lambda}_k,
\end{equation}
$\overline{\lambda}_k=\lambda_k e^{-(\sigma^2_\psi+\sigma^2_\phi)|t-t_0|}$, $\mu_k=({N}-K)\left(\left(1-\frac{1}{N_o}\right)\left(1-\epsilon\right)+1-\lambda_k\right)$, $\xi_k=\beta(\xi^{\rm DL}/(\beta_k P_T))$, and $\beta=K/N>0$. Here, $a_k$ and $c_k$ represent the multiuser interference received at the $k^{\rm th}$ MT
and the variance of the gain of the desired signal, respectively.
%%%%%%%%%%%%%%%%%%%%%%%%%%%%%%%%%%%%%%%%%%%%%%%%%%%%%%%%%%%%%%%%%%%%%%

\subsection{Upper Bound on the Eavesdropper's Capacity}\label{s3c}
%%%%%%%%%%%%%%%%%%%%%%%%%%%%%%%%%%%%%%%%%%%%%%%%%%%%%%%%%%%%%%%%%%%%%%
Following \cite[Theorem 2]{zhu}, a tight and tractable upper bound on eavesdropper's capacity is given by
\begin{equation}\label{Cup1}
C_E \leq \overline{C}_E=\log_2 \left(1+ \frac{p N_E}{q (L - N_E)}\right),
\end{equation}
for $N \to \infty$ and $L  > N_E$.

%\textit{Remark 1}: The number of LOs, $N_o$, affects the ergodic secrecy rate via the terms $a_k$, $c_k$, and $\mu_k$ in the achievable ergodic rate in (\ref{Rkunder}). For $N\to\infty$, $a_k$ and $c_k$ are decreasing
%functions of $N_o$, i.e., less multiple access interference is caused if fewer LOs are employed, whereas $\mu_k$ is an increasing function in $N_o$. Therefore, considering the specific form of the
%denominator of the fraction inside the logarithm in (\ref{Rkunder}), the optimal value of $N_o$ as far as the ergodic secrecy rate is concerned depends on $\phi$.
%%%%%%%%%%%%%%%%%%%%%%%%%%%%%%%%%%%%%%%%%%%%%%%%%%%%%%%%%%%%%%%%%%%%%%
%%%%%%%%%%%%%%%%%%%%%%%%%%%%%%%%%%%%%%%%%%%%%%%%%%%%%%%%%%%%%%%%%%%%%%
\section{Numerical Examples}\label{s5}
%%%%%%%%%%%%%%%%%%%%%%%%%%%%%%%%%%%%%%%%%%%%%%%%%%%%%%%%%%%%%%%%%%%%%%
In this section, we provide numerical and simulation results to verify the analysis presented in Section \ref{s3} and to illustrate the impact of phase noise on the ergodic secrecy rate. For the numerical results,
we numerically evaluate the analytical expression for the lower bound on the ergodic secrecy rate obtained by combining (\ref{rseck}), (\ref{Rkunder}), and (\ref{Cup1}). For the simulation results, we evaluate (\ref{rseck}) using
$\underline{R}_k^{\rm sec}(t) = \log_2(1+\gamma_k(t))$ and $C_E=\mathbb{E}[\log_2(1+\gamma_E)]$ with $\gamma_k(t)$ and $\gamma_E$ given by (\ref{gammak}) and (\ref{CE}), respectively, by Monte Carlo simulation based on
$5,000$ independent channel realizations. In this paper, we adopt spatially orthogonal pilot sequences of length $B=K$ \cite{zhu3}. An example for this type of pilot sequences is provided in \cite[Eq. (7)]{MRC}. For simplicity, in this section, we assume the path-losses for all MTs are identical, i.e., $\beta_k=1$, $1 \leq k \leq K$, and the coherence block length
is equal to $T=500$ time slots. Typical values for the phase noise increment standard deviations, $\sigma_\psi$ ,$\sigma_\phi$, used include $0.06^{\circ}$,  which was adopted in  the long-term evolution
(LTE) specifications \cite{LTE}, and $6^{\circ}$, which corresponds to strong phase noise according to \cite{strongpn1}. The specific values of other adopted system parameters are provided in the captions of the figures.
%%%%%%%%%%%%%%%%%%%%%%%%%%%%%%%%%%%%%%%%%%%%%%%%%%%%%%%%%%%%%%%%%%%%%%
%%%%%%%%%%%%%%%%%%%%%%%%%%%%%%%%%%%%%%%%%%%%%%%%%%%%%%%%%%%%%%%%%%%%%%

\begin{figure}
 \centering
\includegraphics[width=2.8in]{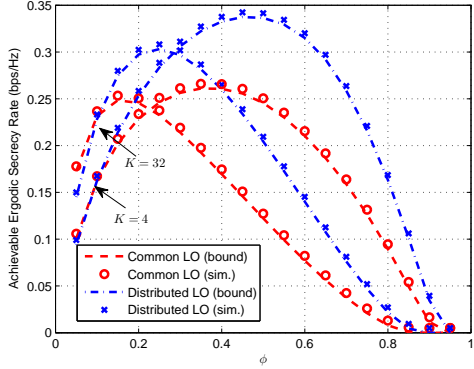}\\
    \caption{Achievable ergodic secrecy rate vs.~$\phi$ for a system with $N=128$, $\sigma_\psi=\sigma_\phi=6^\circ$, $N_E=4$, $p_\tau=P_T/K$, and $P_T=10$ dB.}\label{fig1}
\end{figure}

Fig.~\ref{fig1} shows the achievable ergodic secrecy rate as a function of the power allocation parameter $\phi$ for different numbers of deployed LOs ($N_o=1,128$) and different numbers of MTs ($K=4,32$) for the case of strong phase noise ($\sigma_\phi=\sigma_\psi=6^\circ$). We observe that distributed LOs ($N_o=N$) outperform a common LO ($N_o=1$) for most values of $\phi$, because the desired signal power is independent of $N_o$, while the multiuser interference power is reduced when distributed LOs are deployed, cf. Lemma 2 and Remark 1. This is in accordance with the observation made in \cite{MRC} for the MTs' achievable rate. On the other hand, for small values of $\phi$, a common LO is able to achieve a similar (or even slightly higher) secrecy rate as distributed LOs. This is because the AN leakage, represented by $\mu_k$ in (\ref{Rkunder}), increases in $N_o$, and becomes dominant when a large amount of power is assigned for AN emission, cf. Remark 2.

%Nevertheless, when the optimal $\phi^*$ is adopted, distributed LOs are always preferable to achieve a better secrecy rate, at the expense of increasing the deployment cost by having $N$ LOs.

\begin{figure}
 \centering
\includegraphics[width=2.8in]{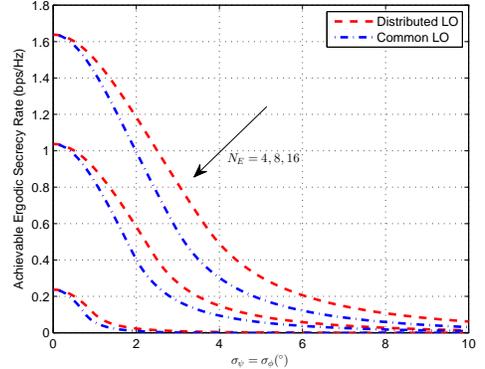}\\
    \caption{Achievable ergodic secrecy rate vs. phase noise standard deviation $\sigma_\psi=\sigma_\phi$ for a system with $K=4$ , $N=128$, $p_\tau=P_T/K$, and $P_T=10$ dB. The optimal $\phi=\phi^*$ maximizing the secrecy rate is adopted.}\label{fig2}
\end{figure}

Fig.~\ref{fig2} depicts the achievable ergodic secrecy rate as a function of the phase noise standard deviation $\sigma_\psi=\sigma_\phi$ for different numbers of eavesdropper antennas. The cases of a common LO and distributed LOs are considered. As expected, the secrecy rate is monotonically decreasing in the phase noise variance as well as in $N_E$. Strong phase noise degrades the MTs' achievable rate, while a larger number of eavesdropper antennas improve Eve's capacity. Both effects are detrimental to the achievable secrecy rate. Furthermore, this figure also suggests that distributed LOs achieve a higher secrecy rate compared to a common LO when the optimal $\phi^*$ maximizing the secrecy rate is employed.
\vspace{-2mm}
%%%%%%%%%%%%%%%%%%%%%%%%%%%%%%%%%%%%%%%%%%%%%%%%%
\section{Conclusions}\label{s6}
%%%%%%%%%%%%%%%%%%%%%%%%%%%%%%%%%%%%%%%%%%%%%%%%%
In this paper, we have investigated the impact of multiplicative phase noise on the secrecy performance of downlink massive MIMO systems employing MF data precoding and NS AN precoding at the BS. For the considered system, a closed-form lower bound
on the achievable ergodic secrecy rate of the users was derived. This bound can be used to obtain insights regarding the impact of system parameters, such as the number of LOs and the phase noise variances, on system performance. Our analytical and simulation results revealed that distributively deployed LOs can achieve a better performance than one common LO as long as the power assigned for data transmission is sufficiently large.
%%%%%%%%%%%%%%%%%%%%%%%%%%%%%%%%%%%%%%%%%%%%%%%%%%%%%%%%%%%%%%%%%%%%%%
%%%%%%%%%%%%%%%%%%%%%%%%%%%%%%%%%%%%%%%%%%%%%%%%%%%%%%%%%%%%%%%%%%%%%%

%%%%%%%%%%%%%%%%%%%%%%%%%%%%%%%%%%%%%%%%%%%%%%%%%%%%%%%%%%%%%%%%%%%%%%

\end{document}